\documentclass[aps,pra,twocolumn,showpacs,superscriptaddress]{revtex4-1}

\usepackage{amsfonts,mathrsfs,amsmath,amsthm,amssymb}
\usepackage{graphicx}
\usepackage{dcolumn}
\usepackage{bm,times}
\usepackage[colorlinks=true,breaklinks=true,linkcolor=blue,citecolor=blue,urlcolor=blue]{hyperref}

\def \tr{ {\rm{Tr}}}
\def \non {\nonumber}
\def \ket {\rangle}
\def \bra {\langle}

\newcommand{\be}{\begin{eqnarray}}
\newcommand{\ee}{\end{eqnarray}}

\newtheorem{lemma}{Lemma}

\makeatletter
    
    \newcommand{\Rmnum}[1]{\expandafter\@slowromancap\romannumeral #1@}
\makeatother

\begin{document}

\title{Geometric discord: A resource for increments of quantum key distribution through twirling}

\author{Xiaohua Wu}
\email{wxhscu@scu.edu.cn}
\affiliation{College of Physical Science and Technology, Sichuan University, 610064, Chengdu, China}

\author{Tao Zhou}
\email{taozhou@swjtu.edu.cn}
\affiliation{Quantum Optoelectronics Laboratory, School of Physical Science and Technology, Southwest Jiaotong University, Chengdu 610031, China}

\date{\today}

\begin{abstract}
In the present work, we consider a scenario  where an arbitrary two-qubit pure state is applied to generate a randomly distributed key via the generalized EPR protocol. Using the twirling  procedure to convert  the pure state into a Werner state, the  error rate of the key can be reduced by a factor of $2/3$. This effect indicates that entanglement is not the sufficient resource of the generalized EPR protocol  since it is not increased in the twirling procedure. Instead of entanglement, the geometric discord is suggested to be the general quantum resource for this task.
\end{abstract}
 \maketitle

\textit{Introduction.---}
How to quantify and characterize the nature of correlations in a quantum state, has a crucial applicative importance in the field of quantum information processing~\cite{Nielsenbook} beyond the fundamental scientific interest. It is well known that a bipartite quantum state can contain both classical and quantum correlations. Quite recently, quantum discord was introduced as a more general measure of quantum correlation~\cite{Ollivier,Henderson} beyond the quantum entanglement~\cite{RMP.81.865}. Since it was regarded as a resource for quantum computation~\cite{Datta}, quantum state merging~\cite{Madhok,Cavalcanti}, and remote state preparation~\cite{Dakic1}, quantum discord has attracted much attention in recent works~\cite{Datta,Madhok,Cavalcanti,Dakic1,Luo,Luo2,Ali,Modi,Dakic2,Bellomo,discord}.

  Among all the known quantum tasks, quantum key distribution (QKD) is one of the most important cases that have been widely discussed in both the theoretic and experimental aspects
  ~\cite{Gisin}. It is well known that the maximally entangled states, or the EPR pairs, can be used to complete the QKD task via the EPR protocol~\cite {Nielsenbook}. Different from the BB84 protocol where the key is transferred from Alice to Bob ~\cite{BB84}, the key  is generated in the EPR scheme:  It is undetermined until Alice or Bob performs a measurement on their EPR parts, respectively.

In the present work, we develop a generalized EPR protocol where an arbitrary two-qubit state is applied to generate a randomly distributed key. The error rate of the generated key can be taken as the figure of merit for this task. A pure state can be converted into a Werner state in a twirling procedure, and the  error rate of the key can be reduced by a factor of $2/3$. It has already been known that twirling can never increase the entanglement, and therefore, the observed effect, where twirling effectively improves the performance of the pure state in the generalized EPR protocol, shows that entanglement is not the sufficient resource for this task. Instead, the geometric discord can be increased in the twirling procedure, and we may conclude that the geometric discord may be the quantum resource in the generalized EPR protocol. Furthermore, with the careful analysis for the general two-qubit case,  we deduce the relation between the error rate and the geometric discord. Based on this, the observed effect may be well explained by the fact that the geometric discord of the pure state can indeed be increased by twirling.

\textit{The general EPR protocol for QKD.---}
To process on, we should first notice that an arbitrary two-qubit state $\rho$ can always be expressed as
\begin{equation}
\label{density}
\rho=\frac{1}{4}\big(\mathbb{I}\otimes \mathbb{I}+x_3\sigma_3\otimes \mathbb{I}+y_3 \mathbb{I}\otimes \sigma_3+\sum_{i,j=1}^3T_{ij}\sigma_i\otimes \sigma_j\big)
\end{equation}
in a fixed basis carefully chosen, where $\sigma_1=|\uparrow\ket\bra\downarrow|+|\downarrow\ket\bra\uparrow\vert$, $\sigma_2=-i|\uparrow\ket\bra\downarrow|+i|\downarrow\ket\bra\uparrow|$, and $\sigma_3=|\uparrow\ket\bra\uparrow|-|\downarrow\ket\bra\downarrow|$ are the Pauli operators, and $T_{ij}=\tr[\rho(\sigma_i\otimes\sigma_j)]$. Assume that the states above consist of two spin-$1/2$ particles labeled by $1$ and $2$, and Alice measures particle $1$ with a fixed observable $\sigma_{\bm a}=\bm\sigma\cdot\bm a$, while Bob performs a measurement on the particle $2$ with the observable $\sigma_{\bm b}=\bm\sigma\cdot\bm b$, where $\bm a$ and $\bm b$ are two unit vectors. Then, \emph{a joint measurement for the observable $\sigma_{\bm a}\otimes\sigma_{\bm b}$ is called to be optimal if and only if $\tr[\rho(\sigma_{\bm a}\otimes\sigma_{\bm b})]=\max_{\bm n}\bra\sigma_{\bm a}\otimes\sigma_{\bm n}\ket$.} For simplicity, hereafter, we denote $\bra\bm a\otimes \bm b\ket=\bra\sigma_{\bm a}\otimes\sigma_{\bm n}\ket$. Now four probabilities $\omega_{\pm\pm}(\bm a,\bm b)$ can be introduced, \emph{i.e.}, $\omega_{++}(\bm a,\bm b)$ is the corresponding probability in the case that the measurement results for both particles are positive, when the joint measurement $\sigma_{\bm a}\otimes\sigma_{\bm b}$ has been performed.  Then, for an arbitrary two-bit state $\rho$, one should have
\be
\omega_{++}(\bm a,\bm b)+\omega_{+-}(\bm a,\bm b)+\omega_{-+}(\bm a,\bm b)+\omega_{--}(\bm a,\bm b)=1,\non
\ee
and the correlation function, $\bra\bm a\otimes \bm b\ket$,  can be expressed as
\be
\bra\bm{a}\otimes \bm{b}\ket&=&\omega_{++}(\bm{a},\bm{b})+\omega_{--}(\bm{a},\bm{b})\non\\
&&-\omega_{+-}(\bm{a},\bm{b})-\omega_{-+}(\bm{a},\bm{b}).\non
\ee
With the optimal measurement defined  above,  the maximally entangled states are the ones satisfying $\bra\bm a\otimes \bm b\ket_\mathrm{max}=1$ for an arbitrary vector $\bm a$.

Now, we come to the EPR protocol for QKD. It is well known that maximally entangled states can be applied to generate a randomly distributed key as in the following arguments~\cite{Nielsenbook}: 

(i) A large amount of EPR pairs shared by Alice and Bob are prepared, and  Alice (Bob) randomly measures her (his) particle of a EPR pair with $\sigma_{\bm a}$ or $\sigma_{\bm a'}$ ($\sigma_{\bm b}$ or $\sigma_{\bm b'})$, where $\bm a'\bot\bm a$, $\bm b'\bot\bm b$~\cite{explain};

(ii) After sufficient runs of measurements have been performed, Alice and Bob exchange the information about the observable used in each run over a public channel;

(iii) The experimental data from the measurements for the obserbles $\sigma_{\bm a}\otimes\sigma_{\bm b'}$ and $\sigma_{\bm a'}\otimes\sigma_{\bm b}$ are discarded. In other words, the remaining data come from the measurements performed by the observbles $\sigma_{\bm a}\otimes\sigma_{\bm b}$ and $\sigma_{\bm a'}\otimes\sigma_{\bm b'}$;

(iv) Finally, by arranging their own remaining experiment data in time sequence, each observer can obtain a random key, a long string of symbols like ``$++-+-\cdots+$''.

The QKD task realized in this way is usually called the EPR protocol since the maximally entangled states (EPR pairs) are used in this procedure. Furthermore, the EPR protocol above can be modified to a more general scenario.

In the general EPR protocol, the EPR pair are replaced by the states in Eq.~(\ref{density}), and the differences come from the following two aspects: 

(i) To get a random distributed key, it is necessary that the two eigenvectors of $\sigma_{\bm a}$ ($\sigma_{\bm a'}$) should appear with equal probability in each measurement. For the state $\rho$ in Eq.~(\ref{density}), it is required that $\bm a$ ($\bm a'$) should be chosen in the $x-y$ plane of the Bloch sphere. For simplicity, we choose that $\bm a=\bm x=(1,0,0)$ and $\bm a'=\bm y=(0,1,0)$.

(ii) The keys in Alice's site may be different from the ones in Bob's site, and the following two measurable quantities, $\delta^{\bm x}(\rho)=\omega_{+-}(\bm x,\bm b)+\omega_{-+}(\bm x,\bm b)$, and $\delta^{\bm y}(\rho)=\omega_{+-}(\bm y,\bm b')+\omega_{-+}(\bm y,\bm b')$, can be used to characterize the discrepancy.
The physical meaning of $\delta^{\bm x}$ and $\delta^{\bm y}$ is clear: They are the probabilities that Alice's measurement result is different from the one of Bob's when the joint measurement $\sigma_{\bm x}\otimes\sigma_{\bm b}$ and $\sigma_{\bm y}\otimes\sigma_{\bm b'}$ are performed, respectively.

Based on the condition that Alice (Bob) selects $\bm x$ and $\bm y$ ( $\bm b$ and $\bm b'$) with equal probability, it is reasonable to define \emph{the (average) error rate of the key}, to be
\be
\delta(\rho)=\frac{1}{2}\big[\delta^x(\rho)+\delta^y(\rho)\big],\non
\ee
and it can be taken as the figure of merit to quantify the general EPR protocol designed above. With the two equalities, $\delta^{\bm x}(\rho)=(1-\bra\bm x\otimes \bm b\ket)/2$ and $\delta^{\bm y}(\rho)=(1-\langle\bm y\otimes \bm b'\ket)/2$, one can obtain
\be
\label{errorate}
\delta(\rho)=\frac{1}{2}-\frac{1}{4}(\bra\bm x\otimes \bm b\ket+\bra\bm y\otimes \bm b'\ket),
\ee
which shows that the error rate is decided by the expectation values of the two observables $\sigma_{\bm x}\otimes\sigma_{\bm b}$ and $\sigma_{\bm y}\otimes \sigma_{\bm b'}$ introduced before.

\textit{Twirling and its effects.---}
In 1989, Werner gave a one parameter family of twirling invariant states which do not violate the Bell inequality although these states are entangled~\cite{Werner}. Since then, twirling has been widely discussed in many quantum tasks, such as the entanglement distillation ~\cite{Bennett,Horodecki} and quantum process tomography ~\cite{tomography}. Following the definition in Ref.~\cite{Horodecki}, any two-qubit state $\rho$ subjected to the $U\otimes U^*$ twirling, can produce a Werner state $\rho_{W}(F)$ as
\be
\label{twirling}
\rho_{W}(F)=\mathcal{T}(\rho)\equiv\int_{U\in \mathrm{SU}(2)}U\otimes U^*\rho(U\otimes U^*)^{\dagger}\d U
\ee
  with $F=\bra\Phi^\pm\vert\rho\vert\Phi^+\ket$, and the maximally entangled state $|\Phi^\pm\ket=(|\uparrow\uparrow\ket\pm|\downarrow\downarrow\ket)/\sqrt{2}$. In the present paper, we use the same symbol $\Phi^+$ to denote the density operator of the pure state, say $\Phi^+=|\Phi^+\ket\bra\Phi^+|$, if no confusion is caused. With $|\Psi^{\pm}\ket=(|\uparrow\downarrow\ket\pm|\downarrow\uparrow\ket)/\sqrt{2}$,
a Werner state $\rho_{W}(F)$ in Eq.~(\ref{twirling}) is
\be
\label{wernerstate}
\rho_{W}(F)=F\Phi^{+}+\frac{1-F}{3}(\Phi^-+\Psi^++\Psi^-),
\ee
where $F$ is a real number, and $ 0\leq F\leq 1$. For the two-qubit states, the Werner states are the unique ones which are invariant under the twirling procedure~\cite {Werner}.

With certain local unitary transformations,  a bipartite pure state can always be expressed as
\be
\label{bistate}
\vert\Omega\rangle=\cos(\frac{\pi}{4}-\frac{\gamma}{2})\vert\uparrow\uparrow\rangle
+\sin(\frac{\pi}{4}-\frac{\gamma}{2})\vert\downarrow\downarrow\rangle,
\ee
with $\gamma$ a free parameter, $0\leqslant\gamma\leqslant \pi/2$. When $\gamma=\pi/2$, $\vert\Omega\rangle$ is a product state, $\vert\Omega\rangle=\vert\uparrow\uparrow\rangle$.  From the definition in Eq.~(\ref{twirling}), it is easy to verify that the pure state in Eq.~(\ref{bistate}) subjected to the twirling can produce a Werner state with $F=\cos^2(\gamma/2)$.

In the  QKD task developed  in following argument, an arbitrary two-qubit state is applied to generate a randomly  distributed key, there exist some cases where  twirling may reduce the error rate of the key. As an important example,   by performing twirling on  the pure state,  the  error rate of the key  will be effectively reduced,
\be
\label{relation}
\delta(\mathcal{T}(\Omega))=\frac{2}{3}\delta(\Omega).
\ee
The derivation of this equation is in the following.

\textit{Reducing the error rate by twirling.---}
As we have shown in Eq.~(\ref{relation}),  the error rate of the key generated with the pure state can be effectively reduced by twirling. An analysis for this effect can be given here. First, for the state in Eq.~(\ref{density}), by some algebra, one can obtain
\be
\label{optimal}
\bra\bm x\otimes \bm b\ket_\mathrm{max}=\sqrt{\sum_{j=1}^3T_{1j}^2},\ \
\bra\bm y\otimes \bm b'\ket_\mathrm{max}=\sqrt{\sum_{j=1}^3T_{2j}^2}.
\ee
Then, for the pure state in Eq.~(\ref{bistate}), the density operator can be written as
\be
\Omega(\gamma)&=&\frac{1}{4}\big[\mathbb{I}\otimes \mathbb{I}+\sin\gamma(\sigma_3\otimes \mathbb{I}+\mathbb{I}\otimes\sigma_3)\non\\
&&+\cos\gamma(\sigma_1\otimes\sigma_1-\sigma_2\otimes\sigma_2)+\sigma_3\otimes\sigma_3\big].\non
\ee
and therefore, with the optimal settings $\bm b=(1,0,0)$ and $\bm b'=(0,-1,0)$, we can obtain $\bra\bm x\otimes \bm b\ket_\mathrm{max}=\bra\bm y\otimes \bm b'\ket_\mathrm{max}=\cos\gamma$. The minimum error rate $\delta(\Omega)=\sin^2(\gamma/2)$.

Meanwhile, the Werner state in Eq.~(\ref{wernerstate}) has an equivalent form, 
\be
\rho_{W}(F)=\frac{1}{4}[\mathbb{I}\otimes\mathbb{I}+\frac{4F-1}{3}(\sigma_1\otimes\sigma_1-\sigma_2\otimes\sigma_2+\sigma_3\otimes\sigma_3)],\non
\ee
and with the same optimal settings as the pure state, we have  $\bra\bm x\otimes \bm b\ket_\mathrm{max}=(4F-1)/3$. By taking $F=\cos^2(\gamma/2)$, the minimum error rate of the pure state after twirling is $\delta(\mathcal{T}(\Omega))=\frac{2}{3}\sin^2(\gamma/2)$, which exactly gives the result in Eq.~(\ref{relation}).

\textit{Geometric discord as a resource for QKD.---}
It has been mentioned before that the effect of twirling shown in Eq.~(\ref{relation}) indicates that entanglement is not the sufficient resource to realize the general EPR protocol, and hence some other quantum resource beyond entanglement should be responsible for this. In the present work, we argue that quantum geometric discord may be viewed as this kind of quantum resource. Our argument is based on the following two aspects. 

(i) For the general two-qubit states, there exists a relation between the minimum error rate and the geometric discord, or more specifically, we have the following lemma.
\begin{lemma}
The geometric discord for a general two-qubit state, $\mathcal{D}_{\mathrm{g}}(\rho)$, is bounded by two optimal values $\delta^{\bm x}_\mathrm{min}(\rho)$ and $\delta^{\bm y}_\mathrm{min}(\rho)$ such that
\be
\label{bound}
\mathcal{D}_{\mathrm{g}}(\rho)\leqslant\big[\frac{1}{2}-\delta^{\bm x}_\mathrm{min}(\rho)\big]^2+\big[\frac{1}{2}-\delta^{\bm y}_\mathrm{min}(\rho)\big]^2.
\ee
\end{lemma}
{\it Proof:} To verify this relation, we should recall the definition of the geometric discord ss the first step. If Alice performs  an arbitrary projective measurement $\{\Pi^a_i\}$ on $\rho$, the final state of the joint system is $\chi_{\rho}=\sum_i\Pi^a_i\otimes\mathbb{I}\rho\Pi^a_i\otimes\mathbb{I}$. Usually, $\chi_{\rho}$ is regarded as  the classic-quantum (CQ) state. With the squared Hilbert-Schmidt norm, $||A||^2=\tr(AA^{\dagger})$, the geometric discord is defined as $\mathcal{D}_{\mathrm{g}}(\rho)=\mathrm{min}_{\Pi^a}||\rho-\chi_\rho||^2$~\cite{Dakic2}. Following the result in Ref.~\cite{Bellomo}, this quantity can also be expressed as the difference of two purities,
\be
\label{geodis}
\mathcal{D}_{\mathrm{g}}(\rho)=\vert\vert \rho\vert\vert^2-\max_{\Pi^a}\vert\vert\chi_{\rho}\vert\vert^2.
\ee
Now, we introduce a special CQ-state $\tilde{\chi}_{\rho}$,
\begin{equation}
\label{CQstates}
\tilde{\chi}_{\rho}=\frac{1}{4}(\mathbb{I}\otimes\mathbb{I}+x_3\sigma_3\otimes \mathbb{I}+y_3 \mathbb{I}\otimes \sigma_3+\sum_{j=1}^3T_{3j}\sigma_3\otimes \sigma_j),
\end{equation}
and obviously, this is the final state after that the projective measurement ($\Pi_1=|\uparrow\ket\bra\uparrow|,\Pi_2=|\downarrow\ket\bra\downarrow|$) is performed by Alice. With the definition in Eq.~(\ref{geodis}), one has $\mathcal{D}_{\mathrm{g}}(\rho)\geqslant||\rho||^2-||\tilde{\chi}_{\rho}||^2$. By jointing it with the Eqs. (\ref{errorate},\ref{optimal}) and the relation $4(||\rho||^2-||\tilde{\chi}_{\rho}||^2)=\sum_{j=1}^3T_{1j}^2+\sum_{j=1}^3T_{2j}^2$, the result in Eq.~(\ref{bound}) is easily obtained.\qed

Note that the inequality in Eq.~(\ref{bound}) is saturated if $\tilde{\chi}_{\rho}$ is the closest CQ-state to $\rho$. For the cases where the state $\rho$ has the following two properties: (I) Its closest CQ-state has the form in Eq.~(\ref{CQstates}), and (II) The two correlation functions $\bra\bm x\otimes\bm b\ket$ and $\bra\bm y\otimes \bm b'\ket$ have a same maximum value, say $\bra\bm x\otimes\bm b\ket_\mathrm{max}=\bra\bm y\otimes \bm b'\ket_\mathrm{max}$. Under these conditions,  the relation in Eq.~(\ref{bound})  takes a more compact form,
\begin{equation}
\delta_\mathrm{min}(\rho)=\frac{1}{2}\bigg(1-\sqrt{2\mathcal{D}_{\mathrm{g}}(\rho)}\bigg).
\end{equation}

As a example, we focus on the so-called $X$-type state,
\begin{equation}
\label{Xstate}
\rho_{X}=\left(
            \begin{array}{cccc}
              \rho_{11} & 0 & 0 & \rho_{14}e^{i\gamma_{14}} \\
              0 & \rho_{12} & \rho_{13}e^{i\gamma_{13}} & 0 \\
              0 &\rho_{13}e^{-i\gamma_{13}} & \rho_{33} & 0 \\
              \rho_{14}e^{-i\gamma_{14}} & 0 & 0 & \rho_{44} \\
            \end{array}
          \right),
\end{equation}
where $\rho_{ij} (i,j=1,2,3,4)$ and $\gamma_{ij}$ are real positive numbers. The $X$-states constitute a subclass of the general two-qubit state in Eq.~(\ref{density}) with $ T_{13}=T_{23}=T_{31}=T_{32}=0$. Now, the special CQ-state, $\tilde{\chi}_{\rho}$, should be
\begin{equation}
\label{CQ}
\tilde{\chi}_{\rho}=\frac{1}{4}(\mathbb{I}\otimes\mathbb{I}+x_3\sigma_3\otimes \mathbb{I}+y_3 \mathbb{I}\otimes \sigma_3+T_{33}\sigma_3\otimes \sigma_3).
\end{equation}
As one of the main results given by Bellomo \emph{et. al.}~\cite{Bellomo}, $\tilde{\chi}_{\rho}$ in Eq.~(\ref{CQ}) should be the closest CQ-state to the state $\rho_{X}$ in Eq.~(\ref{Xstate}) if $k_1\leq k_3$, where
\be
\label{quantity}
k_1&=&4(\rho_{14}+\rho_{23}^2),\non\\
k_3&=&2[(\rho_{11}-\rho_{33})^2+(\rho_{22}-\rho_{44})^2].
\ee

(ii) The effect in Eq.~(\ref{relation}) may be well explained by the fact that twirling  increases the geometric  discord of pure state. This result is supported by the following two lemmas.
\begin{lemma} 
For a pure state or a Werner state of a bipartite system, the minimal error rate of the key is
\be
\delta_{\mathrm{min}}=\frac{1}{2}\bigg(1-\sqrt{2\mathcal{D}_{\mathrm{g}}(\rho)}\bigg).\non
\ee
\end{lemma}
{\it Proof:} It is easy to see that both the Werner state in Eq.~(\ref{wernerstate}) and the pure state in Eq.~(\ref{bistate}) belong to the so-called $X$-type states. For the pure state, the quantities in Eq.~(\ref{quantity}) are $k_1=4\cos^2\gamma$ , $k_3=4(1+\sin^2\gamma)$, and $\bra\bm x\otimes \bm b\ket_\mathrm{max}=\bra\bm y\otimes \bm b'\ket_\mathrm{max}=\cos\gamma$, while for the Werner state,
$k_1=k_3=(4F-1)^2/9$, and $\bra\bm x\otimes \bm b\ket_\mathrm{max}=\bra\bm y\otimes \bm b'\ket_\mathrm{max}=(4F-1)/3$. It is obvious that both the pure state and the Werner state satisfy  the conditions (I) and (II) above, which completes the proof.\qed

\begin{lemma}
For a pure state in Eq.~(\ref{bistate}) and the Werner state produced by this state subjected to $U\otimes U^*$ twirling, the entanglement is the same, while the geometric discord is increased.
\end{lemma}

{\it Proof:} It is well known that twirling is an irreversible preprocessing operation, and therefore never increases the entanglement of the state~\cite{Bennett}. To verify that the entanglement of a pure state is unchanged after a twirling procedure, recall that the entanglement of formation (EoF) is a well-defined measure of the entanglement for a two-qubit state $\rho$~\cite{Wootters}
\be
E[\rho]=H_2\bigg(\frac{1+\sqrt{1-\mathcal{C}^2(\rho)}}{2}\bigg),\non
\ee
where $H_2(x)=-x\log_2 x-(1-x)\log_2(1-x)$ is the binary entropy and $\mathcal{C}(\rho)$ is the concurrence of the state $\rho$. Direct calculation shows that, for the pure state in Eq.~(\ref{bistate}) and the Werner state in Eq.~(\ref{wernerstate}), $\mathcal{C}(\Omega(\gamma))=\mathcal{C}(\rho_W(\cos^2\frac{\gamma}{2}))=\cos\gamma$. Therefore, $E[\Omega]=E[\mathcal{T}(\Omega)]$, which means the entanglement is the same.

On the other hand, with Bellomo's result~\cite{Bellomo}, the geometric discord for $X$-state $\rho_X$ is $\mathcal{D}_{\mathrm{g}}(\rho_{X})=2(\rho_{14}^2+\rho_{23}^2)$ for the case $k_1\leqslant k_3$. By some simple algebra, one can obtain 
\be
\mathcal{D}_{\mathrm{g}}(\rho_W)=\frac{1}{2}\bigg(\frac{2\cos\gamma+1}{3}\bigg)^2,\ \
\mathcal{D}_{\mathrm{g}}(\Omega)=\frac{1}{2}\cos^2\gamma.
\ee
It is clear that the twirling operation on the pure state has increased the geometric discord
\be
\label{inequ}
\mathcal{D}_{\mathrm{g}}(\rho_W)\geqslant\mathcal{D}_{\mathrm{g}}(\Omega).
\ee
\qed

\textit{Conclusions and summaries.---}
In the present work, we obtain a simple relation between the error rate of the key and the geometric discord of the shared state in the genera EPR protocol for QKD. It is shown that the minimum error rate of the key can be reduced by a fact of 2/3 in the twirling procedure. One can explain this effect by the increasing of the geometric discord, as the entanglement is kept unchanged, and therefore the geometric discord can be regarded as a general resource in this task.

From the definition in Eq.~(\ref{twirling}), we see that twirling is a series of bi-local operations, and we have shown that twirling can increase the geometric discord of pure states. It should be noticed that this property  of twirling has not been revealed in previous works. For example,   it has been shown that geometric measure of quantumness of multipartite systems with arbitrary dimension cannot increase under any local quantum channel, if the initial state is pure~\cite{Streltsov}. However, as it is shown in Eq.~(\ref{inequ}), the geometric discord is increased when the pure state is subjected to twirling. Besides the pure states, we also find another example, i.e., $\rho=p\Omega+(1-p)/4\mathbb{I}\otimes \mathbb{I} (0<p<1)$,  where the twirling may increase the geometric discord. Actually, under which conditions the twirling may increase the geometric discord of the general  states   is still an open question. We expect that our results could lead to further theoretical or experimental consequences.

\textit{Acknowledgements.---}
The authors are very grateful to Prof. L.-X. Cen for helpful discussions. This work was partially supported by the National Natural Science Foundation of China under the Grant No.~11405136, and the Fundamental Research Funds for the Central Universities of China A0920502051411-56.


\begin{references}
\bibitem{explain} In the standard EPR protocol, $\bm a=\bm b$ and $\bm a'=\bm b'$, while in the following general EPR protocol, this condition is not necessary.
\end{references}
\end{document}